\documentclass[fleqn,10pt]{wlscirep}
\usepackage[utf8]{inputenc}
\usepackage[T1]{fontenc}
\usepackage[numbers]{natbib}
\usepackage[english]{babel}
\usepackage{graphicx,epsfig}
\usepackage{amsmath}
\usepackage{color}
\usepackage{multirow}
\usepackage[normalem]{ulem}
\usepackage{booktabs}
\usepackage{physics}
\usepackage{dsfont}
\usepackage{float}
\usepackage{amsfonts}
\usepackage{eurosym}
\usepackage{url}
\usepackage{xcolor}
\usepackage{siunitx}
\usepackage{comment}

\begin{document}
\title{Quantum-enhanced unsupervised image segmentation for medical images analysis}

\author[1,*]{Laia Domingo}
\author[1]{Mahdi Chehimi}

\affil[1]{Ingenii Inc., New York, USA}

\affil[*]{laia@ingenii.dev}

\date{\today}

\begin{abstract}
Breast cancer remains the leading cause of cancer-related mortality among women worldwide, necessitating the meticulous examination of mammograms by radiologists to identify and characterize abnormal lesions. This manual process demands high accuracy and is often time-consuming, costly, and error-prone. Automated image segmentation using artificial intelligence offers a promising alternative to streamline this workflow. However, most existing methods are supervised, requiring large, expertly annotated datasets that are not always available, and they experience significant generalization issues. Thus, unsupervised learning models can be leveraged for image segmentation, but they come at a cost of reduced accuracy, or require extensive computational resources when working on mammography medical images.  In this paper, we propose the first end-to-end quantum-enhanced framework for unsupervised mammography medical images segmentation that balances between performance accuracy and computational requirements. We first introduce a quantum-inspired image representation that serves as an initial approximation of the segmentation mask. The segmentation task is then formulated as a quadratic unconstrained binary optimization (QUBO) problem, aiming to maximize the contrast between the background and the tumor region while ensuring a cohesive segmentation mask with minimal connected components. We conduct an extensive evaluation of quantum and quantum-inspired methods for image segmentation, demonstrating that quantum annealing and variational quantum circuits achieve performance comparable to classical optimization techniques. Notably, quantum annealing is shown to be an order of magnitude faster than the classical optimization method in our experiments. Our findings demonstrate that this framework achieves performance comparable to state-of-the-art supervised methods, including UNet-based architectures, offering a viable unsupervised alternative for breast cancer image segmentation.
\end{abstract}

\maketitle

\section{Introduction}
\label{sect:intro}
AI-assisted image-based diagnostic is revolutionizing modern healthcare, by supporting doctors in providing more accurate and efficient diagnosis, and personalized treatment options for their patients. Use cases for AI-assisted diagnostics range from diagnosis of cardiovascular diseases~\citep{ahmad2023diagnosis}, neurodegenerative disorders~\citep{hussain2024mind}, cancer screening~\citep{tsietso2023multi}, fracture detection~\citep{hendrix2023musculoskeletal}, and brain tumors segmentation~\citep{jyothi2023deep}. An especially notable application is medical image analysis for breast cancer diagnosis, as breast cancer remains the leading cause of cancer-related deaths among women \citep{siegel2020cancer}. In 2022, breast cancer affected 2.3 million women worldwide, resulting in over 670,000 deaths \citep{who_breast_cancer}. Early detection of tumors is crucial to lowering mortality risk, which can be achieved through regular mammograms and thorough evaluations by expert radiologists \citep{laubysecretan2015}. AI-assisted techniques such as image segmentation, (i.e., the process of identifying areas of interest in mammography images) play a vital role in early diagnosis, by supporting radiologists in extracting key information about the location and size of lesions, enabling faster and more precise diagnoses. 

The core elements of image segmentation involve partitioning an image into meaningful segments by identifying patterns in specific areas. This process requires identifying relevant features to separate different regions within the image. Various algorithms and techniques have been developed to perform this task, with one of the earliest approaches being thresholding \citep{Otsu, shapiro2001computer}, which was originally proposed for binary segmentation tasks. Since then, more advanced methods, such as watershed algorithms \citep{liang2019watershed}, which rely on morphology and total variation models, graph-based segmentation \citep{meng2021graph}, and deep learning models \citep{wang2022medical}, have been widely employed, particularly in medical imaging. These techniques enable precise segmentation of anatomical structures, enhancing diagnostic processes and assisting in treatment planning. For example, convolutional neural networks (CNNs) have been extensively utilized for segmenting medical images by learning spatial hierarchies of features, offering high accuracy in detecting and delineating tissue boundaries~\citep{long2015fully, zhu2018adversarial, xu2022medical}. However, CNNs often experience reduced accuracy due to the loss of spatial resolution throughout the network, which can hinder the detection of irregularly shaped structures. To address this limitation, the UNet architecture \citep{UNET} was introduced, incorporating skip connections that preserve spatial resolution during mask generation. This architecture has demonstrated superior performance in generating more accurate segmentation masks, particularly for breast cancer detection and segmentation \citep{tang2019efficient, li2019cascade, ConnectedUNets, soulami2021breast}.

All of these deep learning models are supervised methods, where annotated samples are required to ensure strong predictive performance in image segmentation. However, supervised deep learning requires high-quality, expertly annotated datasets of medical images, which may not always be accessible due to the scarcity of qualified experts or the high variability of data. This variability can result in models that perform well on training data but struggle to generalize effectively to new cases. To address this issue, unsupervised learning models like thresholding \citep{Otsu, tariq2023multilevel}, unsupervised optimization \citep{gurobi}, k-means~\citep{dhanachandra2015image} and Gaussian Mixture Models~\citep{schwab2024deepgaussianmixturemodel}, work on unlabeled data, identifying patterns and similarities between data points without explicit guidance.  However, due to the absence of labeled datasets, these methods often exhibit lower performance or require significant computational resources. As a result, it is crucial to develop approaches that strike a balance between accuracy and computational efficiency, specifically tailored for breast cancer image analysis.

However, the increasing amount of computational resources demanded by deep learning methods may not be available, due to the challenges in scaling computational resources beyond the limits of classical hardware. Quantum computing is one of the most promising avenues to solve this challenge, due to the promised speedup and predictive performance improvements \citep{PredictiveModels}. Quantum computing capabilities can be used in two ways: (1) \textbf{Quantum or hybrid algorithms}, where quantum hardware is integrated with classical hardware and computational task is assigned to the most appropriate hardware according to computational requirements, and (2) \textbf{Quantum-inspired methods}, where classical algorithms leverage ideas from quantum physics to achieve a computational speedup without requiring dedicated quantum hardware. 

In the field of image segmentation, quantum-inspired methods such as tensor networks were used for the segmentation of brain MR images \citep{Konar2020QI,Konar2023TN}. Additionally, in~\citep{tariq2023multilevel}, the authors utilized a quantum genetic algorithm to determine optimal thresholds for multilevel image segmentation. Furthermore, quantum annealing was leveraged for segmenting radar images \citep{QAradar} and satellite imagery datasets \citep{venkatesh2024q}. While these approaches achieved a good performance when segmenting images where the region of interest (ROI) has a high contrast relative to the background, they require further optimization and enhancements in order to segment mammography images due to the complexity of their segmentation masks. Finally, the work in \citep{wang2024implementation} proposed a hybrid quantum-classical interactive image segmentation technique, where users provide information on the background and ROI to perform segmentation. However, this approach still requires expert input, limiting its potential for fully automated image segmentation. While the existing works presented so far developed quantum-inspired and hybrid quantum-classical models that successfully performed the task of image segmentation while reducing computational resources, all these models failed in achieving a high performance accuracy when operating on the complex mammography medical images, thus failing to satisfy the necessary balance between accuracy and computational complexity.

\begin{figure*}[!ht]
    \centering
    \includegraphics[width=0.8\textwidth]{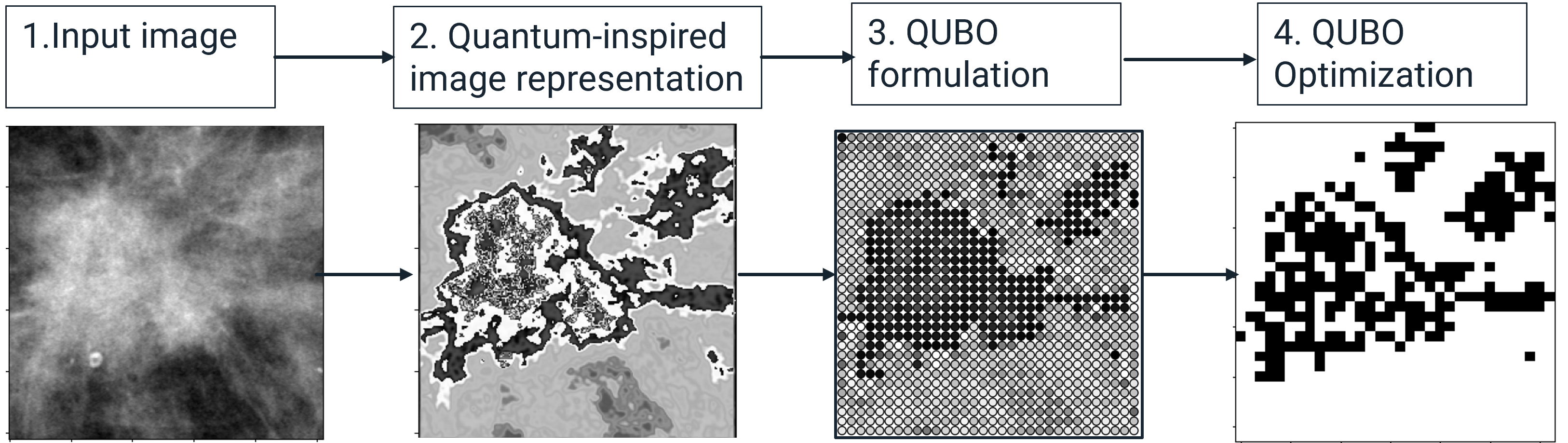}
    \caption{Overview of Quantum Medical Image Classification.}
    \label{fig:overview-approach}
\end{figure*}

The main contribution of this work is the development and assessment of the first end-to-end quantum-enhanced unsupervised pipeline for mammography medical image segmentation that satisfies the intricate balance between performance accuracy and computational complexity. The proposed framework is represented in Fig.~\ref{fig:overview-approach}, which includes: (1) a mammography image to be segmented, (2) a quantum-inspired image transformation, (3) a formal definition of the image segmentation task as a quadratic unconstrained binary optimization (QUBO) problem, and (4) a comparative evaluation of various quantum, quantum-inspired, and classical methods. Specifically, we examine: a) two quantum-inspired techniques, simulated annealing and variational quantum circuits, b) a fully quantum technique, quantum annealing, and c) four classical methods, including two supervised neural network models (UNet and ResUNet), Gurobi optimization, and Otsu thresholding. The results show that the performance of our unsupervised pipeline is comparable to that of the supervised UNet model, and significantly outperforms the traditional Otsu thresholding approach. Additionally, the QUBO problem-solving process is approximately 10 times faster than the classical Gurobi solver for this task, with much lower variability in execution time.

\section{Results}\label{sect:results}

In this section, we present the results of the different image segmentation methods of the INbreast dataset \citep{Inbreast}, a publicly available collection of high-quality digital mammograms with detailed ROI annotations for breast cancer research. Segmentation masks for each mammogram are predicted using different optimization models, including quantum, quantum-inspired, and classical algorithms. To assess the performance of these masks, we employ two quantitative metrics: the Dice score and the IoU score \citep{ConnectedUNets}. Given the predicted mask $\hat{M}$ and the true mask $M$, the Dice similarity score measures the average intersection of the two masks relative to the sum of their areas:

\begin{equation}
    \text{Dice score}(M, \hat{M}) = \frac{2 \times \text{Area } M \cap \hat{M}}{\text{Area } M + \text{Area } \hat{M}}.
    \label{eq:Dice}
\end{equation}
The second metric, known as the Intersection over Union (IoU) score, calculates the ratio of the intersection area to the union area of the masks:

\begin{equation}
    \text{IoU score}(M, \hat{M}) = \frac{\text{Area } M \cap \hat{M}}{\text{Area } M \cup \hat{M}}.
    \label{eq:IoU}
\end{equation}
High performance in the image segmentation task is indicated when both the Dice score and the IoU score are close to 1.


\subsection{Quantum-inspired image representation}
\label{sect:results-qinsp}
\begin{figure*}[!ht]
    \centering
    \includegraphics[width=0.9\textwidth]{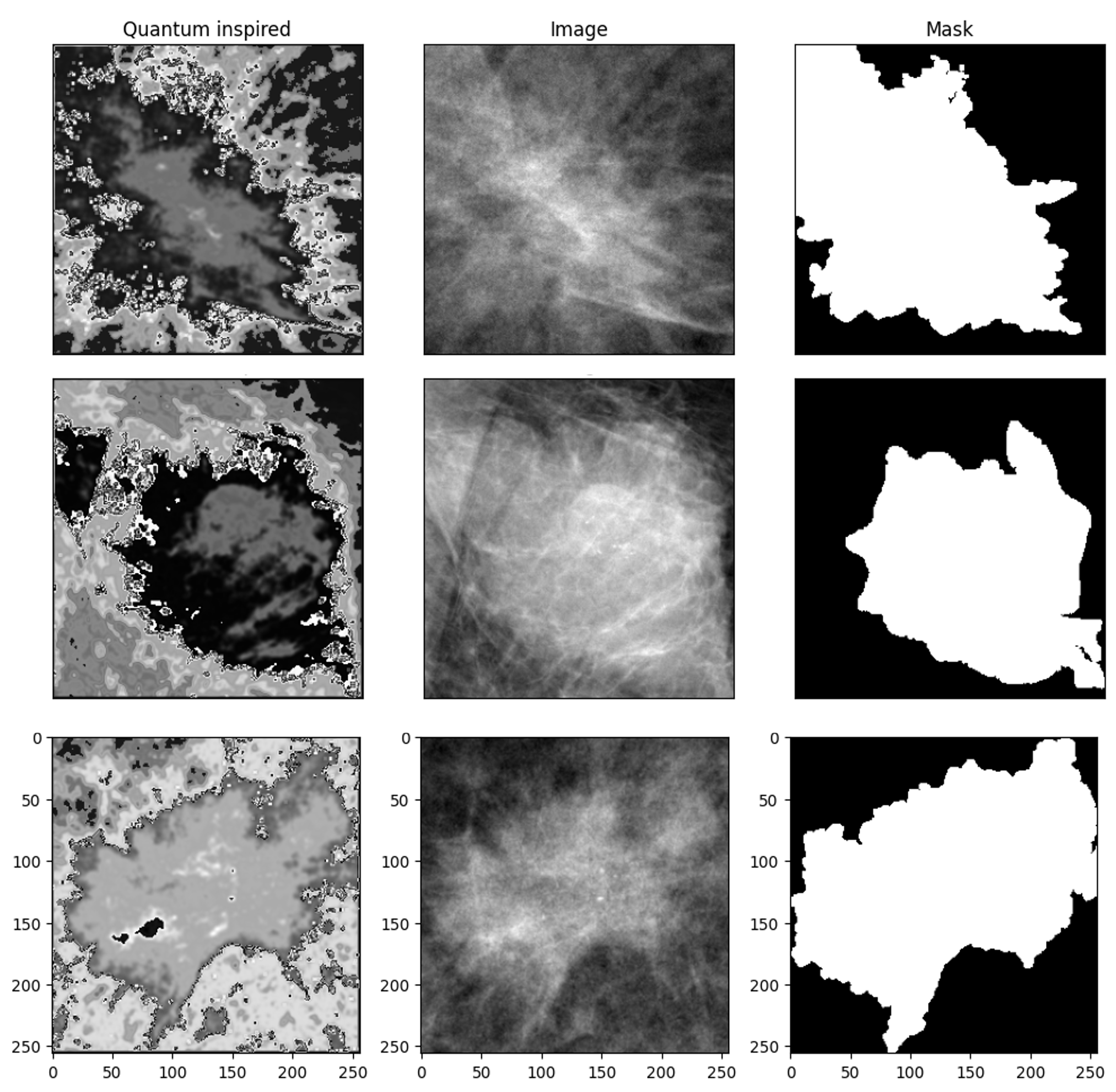}
    \caption{Example quantum-inspired image representations (first column), the original images (second column) and the ROI masks (third column).}
    \label{fig:quantum-filter}
\end{figure*}

In this section, we analyze the performance of the quantum-inspired image transformation presented in Sect. \ref{sect:Image_transformation}, which is designed to highlight regions with varying contrast and identify potential regions of interest. This transformation is used as the first step of the image segmentation pipeline, serving as the input to the optimization algorithms. Figure \ref{fig:quantum-filter} shows three different examples of the quantum-inspired image transformation, alongside the original image and the expected segmentation mask. 
The quantum-inspired transformation effectively delineates the ROI borders and accentuates the varying intensity levels present within the image. This enhancement makes the transformation particularly valuable as a preliminary step in the image segmentation process. By accurately capturing the critical features and boundaries within the image, this approach provides a strong initial approximation that can significantly aid in the subsequent segmentation task, ensuring that the critical areas are appropriately highlighted for further analysis. To further evaluate this hypothesis, we trained two supervised neural network models, specifically the U-Net and ResUNet described in Sect. \ref{sect:classical-methods}, utilizing the quantum-inspired image representation as the input dataset. We then compared the performance of these models against their performance when trained on the original dataset. In order to train the neural network models, we used the following loss function \citep{ConnectedUNets}, which optimizes both the Dice ans IoU scores simultanously:
\begin{eqnarray}
    \mathcal{L}_{UNET}(M, \hat{M}) = - (0.4 \times \text{Dice score}(M, \hat{M}) + 0.6 \times\text{IoU score}(M, \hat{M})).
\end{eqnarray}
The comparative results of the UNET models with both the original dataset and the quantum-inspired representation are presented in Table \ref{tab:unet-results}. 
The convergence of the neural networks is determined using early stopping, ending the training when the validation set loss function no longer decreases for 5 consecutive epochs.
\begin{table}[t]
    \centering
    \caption{Performance of neural network segmentation models on the test set using both the original images and the quantum-inspired image representations, along with the number of training epochs required for the validation loss function to converge.}
    \begin{tabular}{|c|c|c|c|c|}
    \hline
        Model & Input data & Dice score & IoU score & Training epochs\\
        \hline
        UNET & Original & 0.911 & 0.799 & 45\\
        
        UNET & Qnt.insp. & 0.910 & 0.816 & 30 \\
        
        ResUNET & Original & 0.926 & 0.836 & 45\\
        
        ResUNET & Qnt.insp. & 0.947 & 0.857 & 30\\
        \hline
    \end{tabular}
    \label{tab:unet-results}
\end{table}
The results show that the ResUNET performs slightly better than the UNET for both datasets. Additionally, both models demonstrate marginally better performance when trained on the quantum-inspired representations of the mammography images. Notably, the number of epochs required for the loss function on the validation set to converge was reduced by 33\% when using the quantum-inspired images as input. This finding is intuitive, as the quantum-inspired images serve as an initial approximation of the final segmentation mask, making it easier for the U-Net models to learn and accurately segment the images.

While this strategy offers a modest improvement in performance and reduces the number of training epochs compared to the fully classical UNET models, it still relies on a supervised approach, therefore needing an annotated dataset to train the neural networks. Therefore, we explore unsupervised alternatives, using classical, quantum-inspired and fully quantum approaches to generate segmentation masks without using labeled data.

\subsection{Image segmentation}
\label{sect:results-image-segmentation}

In this section, we present a comprehensive comparison of various unsupervised methods for breast cancer image segmentation, utilizing quantum-inspired images as input. The images in these experiments are downsized to $42 \times 42$ pixels to fit quantum requirements, as explained in Sect. \ref{sect:data}. The segmentation problem is formulated as a QUBO problem, where the objective is to minimize the quadratic loss function defined in Eq. \ref{eq:loss}. To address this, we employ a range of quantum and quantum-inspired techniques, including simulated annealing, quantum annealing, and Variational Quantum Algorithms (VQAs), as described in Sect. \ref{sect:Image_segmentation}.

These approaches are benchmarked against two classical unsupervised methods: the Otsu thresholding technique and classical optimization using Gurobi software. Additionally, we include comparisons with two supervised classical models, namely the U-Net and ResUNet architectures. This analysis evaluates the segmentation performance of each method, alongside key considerations such as execution time and scalability.

\subsubsection{Effect of $\alpha$}
\label{sect:alpha}
\begin{figure}[!ht]
    \centering
    \includegraphics[width=0.7\linewidth]{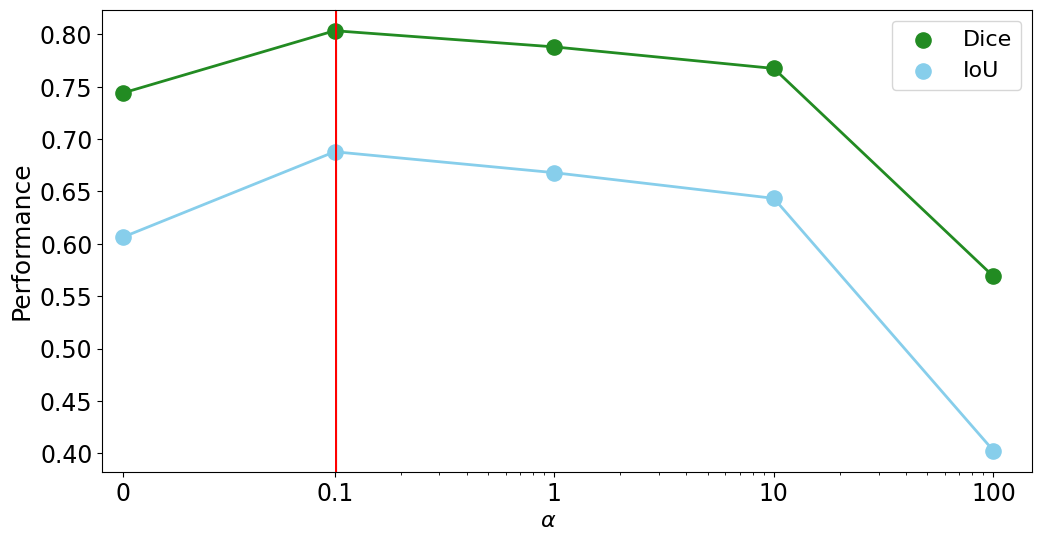}
    \caption{Effect of the value $\alpha$ on the performance of the optimization problem using simulated annealing.}
    \label{fig:alpha}
\end{figure}

The QUBO formulation of the segmentation model is designed to balance two key components: a min-cut term, which promotes maximum dissimilarity between the background and foreground classes, and a smoothness term, which encourages neighboring pixels to belong to the same class. The balance between these two competing objectives is regulated by the hyperparameter $\alpha$. To determine the optimal value of $\alpha$, we conducted experiments by computing the segmented masks using simulated annealing across five different values of $\alpha$—specifically, $0$, $0.1$, $1$, $10$, and $100$—for all the images in the test set. The results of these experiments are visualized in Fig. \ref{fig:alpha}.
The analysis reveals that the best average segmentation performance, as measured by both the Dice coefficient and the Intersection over Union (IoU) score, is achieved when $\alpha = 0.1$. When $\alpha$ is set to low values, the resulting segmentation masks tend to be overly fragmented, with noticeable gaps within the detected regions of interest (ROIs). Conversely, at high values of $\alpha$, the optimization process becomes overly smooth, failing to accurately delineate the edges within the ROI, as it excessively forces neighboring pixels to belong to the same class.

Given these findings, we have chosen to use $\alpha = 0.1$ for the remainder of the experiments conducted in this study, as this value provides the best balance between capturing the necessary detail and maintaining smoothness in the segmentation masks.

\subsubsection{Image segmentation performance}
\label{sect:Performance}
\begin{figure*}[!ht]
    \centering
    \includegraphics[width=0.85\textwidth]{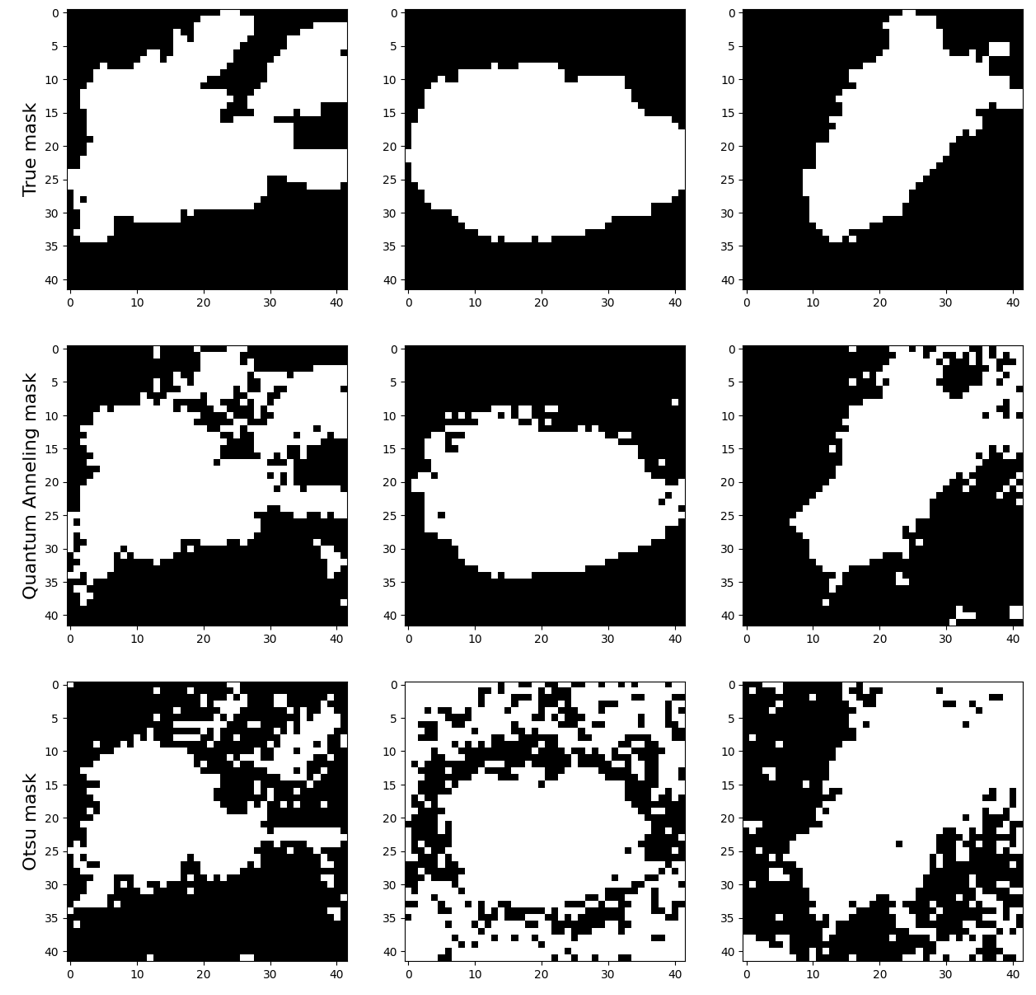}
    \caption{Examples of different image segmentation masks produced by the Otsu thresholding method, the quantum annealing method and the ground truth. }
    \label{fig:masks}
\end{figure*}

Next, we evaluate and compare the image segmentation performance of all quantum methods against classical benchmarks. Table \ref{tab:performance} presents the Dice and IoU coefficients for each method considered in this study. The results demonstrate that all quantum and quantum-inspired methods significantly outperform the Otsu method, a simple classical unsupervised approach commonly used in image processing. Notably, both the quantum annealing and VQA approaches achieve performance nearly identical to that of the classical Gurobi optimization, which we use as the ground truth reference in this context. This high level of performance is closely followed by the simulated annealing method. Interestingly, the best-performing quantum methods exhibit a performance comparable to that of the supervised U-Net model, a state-of-the-art technique in image segmentation, and only slightly below the ResUNet, which is an enhanced version of U-Net. This is particularly noteworthy because these quantum methods operate in an unsupervised manner, without access to annotated datasets for training. Despite this, they offer a highly accurate approximation of the true segmentation masks, highlighting their potential as powerful tools for image segmentation tasks in scenarios where supervised learning is not feasible.
\begin{table}[t]
    \centering
    \begin{tabular}{|c|c|c|c|}
    \hline
        Method & Type & Dice & IoU \\
        \hline
       UNET  & Classical supervised & 0.85 & 0.75\\
       ResUNET & Classical supervised & 0.89 & 0.81 \\
       Otsu & Classical unsupervised & 0.75 & 0.62 \\
       Gurobi & Classical optimization & 0.84 & 0.74 \\
       Simulated annealing & Quantum-inspired & 0.80 & 0.69 \\
       Quantum annealing & Quantum computing & 0.84 & 0.74 \\
       VQA & Quantum-inspired & 0.83 & 0.73\\
       \hline
    \end{tabular}
    \caption{Performance of the image segmentation methods on the $42\times42$ images.}
    \label{tab:performance}
\end{table}
Figure \ref{fig:masks} presents a qualitative comparison of the segmentation masks generated by three different models: Otsu thresholding, the quantum annealing method, and the ground truth. These examples were chosen to illustrate the varying levels of performance across the methods.

The mask produced by the Otsu thresholding method is notably cluttered with background noise, including numerous isolated pixels and even entire regions that do not correspond to the ROI. This excessive noise indicates the method's limitations in accurately segmenting complex images, as it fails to effectively distinguish the ROI from the background.
In contrast, the mask generated by the quantum annealing method bears a closer resemblance to the true segmentation mask. Although it still contains some isolated pixels that could be refined through post-processing techniques, the overall quality of segmentation is significantly higher. The quantum annealing method effectively captures the main structure of the ROI, demonstrating an improvement over the Otsu method. One of the key improvements is that it successfully identifies only one connected component for each ROI, reducing the likelihood of spurious regions that are prevalent in the Otsu segmentation.

\subsubsection{Execution times}
\label{sect:Time}
Lastly, we evaluate the execution times associated with each image segmentation method. One of the most striking observations is that the quantum annealing method, which achieves performance on par with the classical Gurobi optimization, is considerably faster than its classical counterpart. Specifically, the average execution time for quantum annealing is an order of magnitude shorter than Gurobi, and has significantly lower variability in the execution times. The complexity of the Gurobi solver can escalate dramatically, potentially even exponentially, with the number of pixels, depending on the image structure. In our experiments, quantum annealing offers a competing alternative that can significantly reduce computational time for images with thousands of pixels.

The execution time for the quantum annealing method is comparable to that of the Otsu thresholding method, yet it delivers far superior segmentation performance, as discussed earlier. In contrast, the Simulated Annealing approach requires substantially longer execution times than Gurobi optimization, with average optimization times being nearly ten times slower than the classical method. Given that its performance is also inferior to Gurobi, simulated annealing does not present a compelling alternative for image segmentation tasks when compared to classical methods.

The VQA approach, which in this study is implemented as a quantum-inspired method but could eventually be executed on actual quantum hardware, has a computational time that is roughly 100 times longer than that of the Gurobi optimization. However, this extended runtime could be significantly reduced if the algorithm were to be run on real quantum devices, an avenue we plan to explore in future work. Although the current implementation does not offer an execution time improvement over classical software, the VQA method has the potential to excel with large images due to its ability to represent the image using only a logarithmic number of qubits.

Finally, the supervised neural network methods, U-Net and ResUNet, have the longest execution times overall, primarily due to the need for extensive training on annotated datasets. This training phase is computationally intensive and time-consuming, which adds to the overall execution time required for these methods. Despite their strong performance, this extended training time and the need for a training dataset presents a challenge when compared to the unsupervised methods.

\begin{figure}[!ht]
    \centering
    \includegraphics[width=0.7\linewidth]{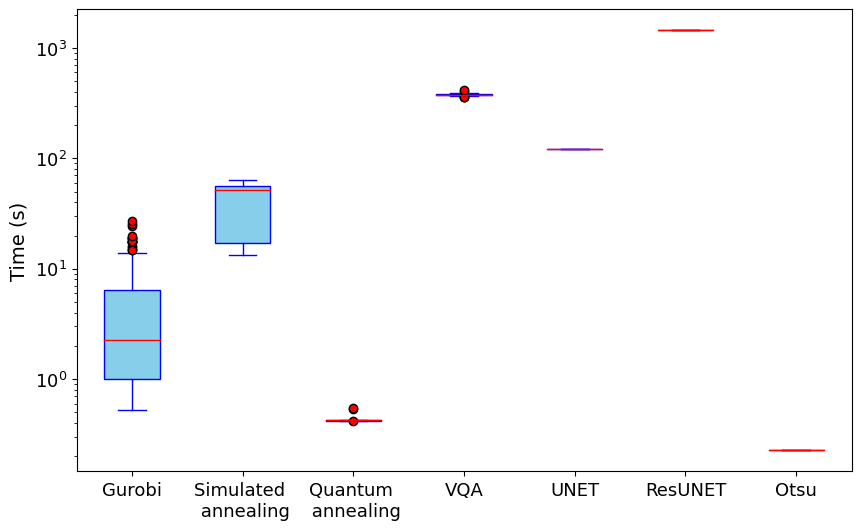}
    \caption{Execution times for the different image segmentation models used in this work.}
    \label{fig:times}
\end{figure}

\section{Conclusion and Future Work}
\label{sect:conclusion}

This study investigates various image segmentation methodologies, including classical, quantum-inspired, and quantum computing approaches, for breast cancer detection using the INbreast dataset. We proposed using a quantum-inspired image transformation technique that effectively enhances the input images by accentuating critical features and boundaries, serving as a powerful pre-processing step for segmentation tasks. This transformation was shown to improve the performance of both supervised and unsupervised segmentation models, with the quantum-inspired representations leading to faster convergence during training and slightly improved segmentation accuracy.

Our comparative analysis revealed that quantum and quantum-inspired methods, particularly quantum annealing and VQAs, achieved performance on par with classical optimization techniques like Gurobi, and even approached the effectiveness of state-of-the-art supervised models such as U-Net and ResUNet. Remarkably, these quantum methods operated in an unsupervised manner, yet they managed to provide a high-fidelity approximation of the true segmentation masks, demonstrating their potential as robust alternatives when labeled data is scarce or unavailable.

The analysis of execution times revealed that quantum annealing outperformed the Gurobi solver by more than an order of magnitude in speed, even for the relatively small case of $42 \times 42$ images. We anticipate that larger quantum computers will provide even greater computational advantages over classical optimization solvers. Additionally, while the current implementation of Variational Quantum Algorithms (VQAs) is constrained by long computational times on classical simulators, this approach demonstrated promising scalability with the number of pixels. These time constraints could be mitigated by deploying VQAs on actual quantum hardware or leveraging efficient quantum circuit simulators, such as tensor networks, which we plan to investigate in future work.

In future works, we plan to extend our investigation to larger datasets and more complex imaging modalities, such as 3D mammography and MRI scans, to further validate the scalability and effectiveness of quantum and quantum-inspired methods. Additionally, we plan to integrate tensor compression techniques using tensor networks \citep{Konar2023QuantumTensor, CompactifAI, CP, tucker, qing2024compressing} to extend this methodology to high-dimensional data without compromising execution times. These efforts will further establish the applicability of quantum-inspired approaches in medical image segmentation.

\section{Methods}
\label{sect:methodology}

In the next sections we outline the methodology employed in this study, detailing (1) the data processing procedures, (2) the quantum-inspired image transformation techniques, and (3) the QUBO formulation of the image segmentation problem, which encompass classical, quantum, and quantum-inspired optimization approaches.

\subsection{Data processing}
\label{sect:data}
In this study, we utilize the INbreast dataset \citep{Inbreast}, a publicly available database of full-field digital mammography images in DICOM format. The dataset comprises 410 mammograms, with only 107 cases featuring mass lesions in both mediolateral oblique (MLO) and craniocaudal (CC) views from 115 unique patients. The raw images were annotated by experts and vary in size, with an average resolution of $3328\times4084$ pixels.

The segmentation models presented in this work are specifically designed to accurately delineate the borders of the segmented mass, which is essential for precise measurement of the mass size and contours. To focus on the relevant areas, the original images were cropped to include only the ROI. This cropping was done by creating boxes around the ROI, which can be detected using machine learning techniques such as You-Only-Look-Once (YOLO) models \citep{ConnectedUNets, YOLO}. YOLO models are suitable for this task because they can be trained on larger datasets without needing highly detailed detection. In this particular study, $256\times256$ pixel boxes around the ROI masses were generated using annotations provided by doctors from the dataset. This pre-processing step ensures that the segmentation model works on a more focused area, improving its accuracy and efficiency.

Since our goal is to evaluate the performance of our unsupervised method versus supervised methods, we apply different types of pre-processing for the input image. For the unsupervised segmentation models employed in this study, no additional preprocessing steps were performed, except for the quantum-inspired image transformation described in the next section. However, for the supervised methods based on neural networks, a data augmentation step was necessary to increase the number of training samples. Specifically, we augmented the data by a factor of 4x by rotating the images by 0º, 90º, 180º, and 270º. Additionally, we further increased the data samples by a factor of 2x by applying Contrast Limited Adaptive Histogram Equalization \citep{Histogram}, which effectively spreads out the most frequent intensity values, thereby enhancing the global contrast of the images. We used a grid size of $8\times8$ and two thresholds for contrast limiting, with values of 1 and 2. 
Using these techniques, we obtained a total of 920 images and their corresponding ROIs. For the supervised methods, the dataset was divided into training, validation, and testing sets, with 15\% of the data allocated for validation and another 15\% for testing. For the unsupervised image segmentation experiments, the images have been downsized to $42 \times 42$ pixels, to fit the requirements of quantum hardware.

\subsection{Quantum-inspired image representation}
\label{sect:Image_transformation}
Ensuring precise image segmentation relies heavily on appropriately representing input images. In this study, we adapt the quantum-inspired method introduced by Konar et al. \citep{Konar2020QI} to achieve optimal representation of quantum images with minimal computational overhead. Given a $N \times M$ mammography image with pixel intensity $I \in \mathbb{R}^{N \times M}$ normalized within the $[0,1]$ range, the quantum filter produces an image $z \in \mathbb{R}^{N \times M}$ according to the following equation:
\begin{equation}
\label{qinspired}
    z_{ij} = \sum_{p,q \in \{-1,0,1\}} \sigma\Big(I_{ij} \braket{\varphi_{pq}^{ij}}{\omega_{ij}}\Big), \quad  i= 1, \cdots N, \ j=1, \cdots M.
\end{equation}
Here, $\ket{\varphi_{pq}^{ij}}$ represents a quantum state encoding the relative intensity difference $\alpha_{pq}^{ij}$ between the pixel under consideration $I_{ij}$ and its neighboring pixel $I_{i+p,j+q}$, given by:
\begin{equation}
    \begin{cases} 
        & \ket{\varphi_{pq}^{ij}} = \cos(\frac{\pi}{2}\alpha_{pq}^{ij}) \ket{0} + \sin(\frac{\pi}{2}\alpha_{pq}^{ij}) \ket{1} \\[8pt]
        & \alpha_{pq}^{ij} = 1 - (I_{i+p,j+q} - I_{ij}).
    \end{cases}
\end{equation}
This relative intensity measure helps to segment the foreground and background regions of an image. Additionally, the quantum state $\ket{\omega_{ij}}$ encodes the contribution $S^{ij}$ of pixel intensities within the $(3 \times 3)$ neighborhood of the pixel $I_{ij}$, defined as:
\begin{equation}
    \begin{cases} 
        & \ket{\omega_{ij}} = \cos(\frac{\pi}{2}S^{ij}) \ket{0} + \sin(\frac{\pi}{2}S^{ij}) \ket{1} \\[8pt]
        & S_{ij} = \sum_{p,q \in \{-1,0,1\}} I_{i+p, j+q}.
    \end{cases}\vspace{0.3cm}
    \label{eq:S}
\end{equation}
Thus, the term $I_{ij}\braket{\varphi_{pq}^{ij}}{\omega_{ij}}$ in Equation~\ref{qinspired} weighs the original pixel intensity by a relative measure of pairwise intensity difference versus total neighborhood contribution. 

Finally, we use a versatile multi-level sigmoid activation function, denoted as $\sigma$, tailored to handle the complexity of multi-level image segmentation tasks \citep{Bhattacharyya2011}. This activation function, a generalized form of the traditional sigmoid function, is adept at producing outputs that correspond to the various gray levels present in an image. As a result, it stands out as a suitable choice for segmenting images with multiple intensity levels. The formulation of this activation function is given by:

\begin{equation}
    \sigma(x) = \frac{1}{\lambda + e^{-\mu(x - \eta)}},
\end{equation}

Here, $\mu$ serves as the steepness factor, functioning as a critical hyperparameter of the model. Higher values of $\mu$ lead to a sharper curve, enhancing the discrimination between different intensity levels, whereas lower values yield a more gradual curve, providing smoother transitions. In our experiments, we fix $\mu$ at a value of $0.4$ to strike a balance between sensitivity and smoothness. The parameter $\eta$, termed the activation parameter, plays a pivotal role in adjusting the midpoint of the sigmoid function's transition. Given the heterogeneous nature of neighborhoods within an image, $\eta$ is adaptively determined for each neighborhood, ensuring flexibility in the segmentation process. Specifically, we set $\eta = S_{ij}$, where $S_{ij}$ represents the cumulative pixel intensities within the local neighborhood centered around pixel $I_{ij}$ (see Eq. \ref{eq:S}). Finally, the parameter $\lambda$ controls the multilevel segmentation classes. Given the number of gray-scale classes $L$, 
\begin{equation}
\label{lambda}
    \lambda = \frac{S_{i,j}}{\omega_{k+1} - \omega_{k}}.
\end{equation}
Here, $\omega = [\omega_1, \cdots, \omega_L]$, with $\omega_1=0$ and $\omega_L = 1$, denotes the distribution of gray-scale contributions across $L$ classes. The choice of $k$ in Equation~\ref{lambda} is determined such that $S_{ij}$ falls within the range defined by $\omega_k$ and $\omega_{k+1}$. By adjusting the spacing between successive values in $\omega$, we can assign varying degrees of importance to different gray-scale levels within the image, thereby enhancing the segmentation process's adaptability and effectiveness. 
\begin{figure}[!ht]
    \centering
    \includegraphics[width=0.6\columnwidth]{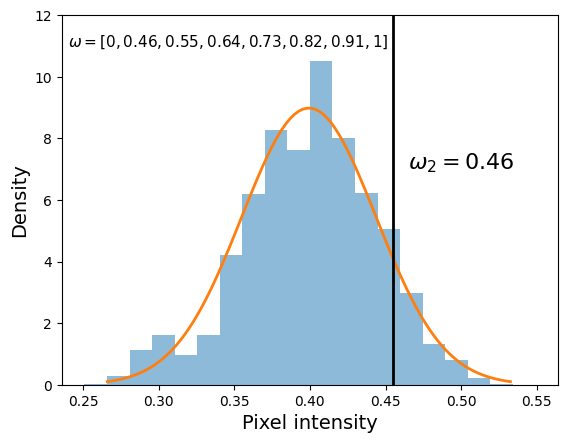}
    \caption{Choice of hyperparameter $\omega$, that controls the gray-scale distribution across the $L$ segmentation classes.}
    \label{fig:weights-quantum-inspired}
\end{figure}
In our study, we fix the number of gray-scale classes to $L=8$ and introduce an adaptive approach for determining the hyperparameter $\omega$. This method allows for enhanced flexibility and effectiveness in image segmentation. Specifically, we designate $\omega_2$ to correspond to the $p$ percentile of the pixel intensity distribution, achieving a balanced emphasis on the brighter areas within the image. Then, $\omega_3, \cdots, \omega_L$ are evenly spaced, so that
\begin{equation}
    \omega = \{0, \omega_2 + k \frac{1 - \omega_2}{L}\}, \quad k \in \{0, \cdots, L\}
\end{equation}
After conducting thorough experimentation with various values of $p$, we opt for $p=0.9$, which effectively accentuates the high-intensity regions, thereby facilitating clearer identification of image edges. An example of the selection of $\omega$ is shown in Fig. \ref{fig:weights-quantum-inspired}.

In contrast to the iterative procedure outlined in the original works by Konar et al. \citep{Konar2020QI, Konar2023TN}, wherein the image transformation process from Equation~\ref{qinspired} is repeated until convergence, we observe that a single iteration of the quantum-inspired transformation suffices to generate a suitable image representation for subsequent predictive modeling tasks. This streamlined approach significantly reduces the computational complexity associated with the quantum-inspired method.


\subsection{Image Segmentation}
\label{sect:Image_segmentation}
The resulting image is then provided as input to an unsupervised segmentation algorithm to identify the ROI within the mammography image. To this end, we model the image segmentation problem as a min-cut problem, following the ideas of Ref. \citep{QA,venkatesh2}, while adding additional constraints to achieve higher-quality segmentation. The first step is to map the processed image to a weighted graph with a grid-based topology. In this graph representation, the nodes correspond to the individual pixels of the image, and the edges portray the similarity between neighboring pixels. Let us denote this graph as $G(V,E,W)$, where $V$ is the set of vertices of the graph, $E \subseteq V \times V$ is the set of edges of the graph, and $W$ is the weight function of the graph, $W: E \mapsto \mathbb{R}$. In this scenario, the weights correspond to the Gaussian similarity of pairs of neighboring pixels. That is, given two pixel intensities $z_i$ and $z_j$, the similarity is defined as:

\begin{equation}
    W_{ij} = \exp\left(-\frac{(z_i - z_j)^2}{2\hat{\sigma}^2}\right),
\end{equation}
where $\hat{\sigma}$ is a hyperparameter of the model. After performing a grid-search parameter optimization, we found that the best balance between the inherent intensity variation and the detection of edges was to select $\hat{\sigma} = 0.5 std(z)$, where $std(z)$ represents the standard deviation of the quantum-inspired image $z$.

\subsubsection{Quadratic Unconstrained Binary Optimization (QUBO)}
QUBO is a mathematical formulation widely used for solving combinatorial optimization problems, such as portfolio optimization, scheduling, and other NP-hard problems \citep{lucas2014ising}. QUBO has become increasingly relevant with the advent of quantum computing technologies, particularly in the context of quantum annealing devices, which are well-suited for solving this type of optimization problem. 
The basic principle of QUBO involves formulating the problem in terms of binary variables and defining a quadratic cost function to minimize. The goal is to adjust these variables iteratively to find the optimal solution \citep{mcgeoch2020theory}. However, despite its potential, current QUBO formulations come with certain limitations. One key constraint is that all variables must be binary, which may require additional transformations for problems that are naturally expressed in other forms. Additionally, the problem must be unconstrained, requiring any additional constraints to be incorporated into the optimization function.

Representing the image using a graph-based structure enables the formulation of the segmentation task as a QUBO problem. Given a binary vector $\vec{x}$, solving a QUBO problem corresponds to minimizing the following expression:
\begin{equation}
    f(\vec{x}) = \vec{x}^T Q \vec{x} + \vec{c}^T \vec{x}.
    \label{eq:QUBO}
\end{equation}
In this work, the optimal image segmentation is  determined by identifying the minimum cut of the graph that satisfies a smoothness constraint, promoting spatial coherence in the segmentation mask. Finding the min-cut corresponds to the partition of the image that minimizes the sum of the weights associated with the cut edges, which guarantees minimum similarity between the two different classes. Given a segmentation mask $A$ and its complementary $\bar{A}$, the min-cut formulation minimizes the following cost function:
\begin{equation}
    \displaystyle \mathcal{L}(G) = \sum_{v_i \in A, v_j \in \bar{A}} W_{ij}.
    \label{eq:classical-mincut}
\end{equation}
This min-cut problem can be formulated as a QUBO problem. Let $\vec{x}$ be a binary variable vector denoting the class of all pixels, such that
\begin{equation}
    x_i = 
        \begin{cases} 
        0 & \text{if } i \in A \\
        1 & \text{if } i \in \bar{A} 
        \end{cases}
    \quad \forall i \in V.
    \label{eq:variables}
\end{equation}
Then, minimizing Eq. \ref{eq:classical-mincut} is equivalent to minimizing
\begin{equation}
    \sum_{(i,j) \in E} W_{ij}|x_i - x_j| = W_{ij}(x_i + x_j -2x_ix_j).
\end{equation}
To improve the QUBO formulation of the min-cut problem for our medical image segmentation task, we incorporate a smoothness penalty than ensures a cohesive image segmentation masks, Following Potts model \citep{Potts}, this constraint ensures that neighboring nodes are more likely to be in the same subset. Thus, the final loss function is
\begin{eqnarray}
    &\mathcal{L}(\vec{x}) = \displaystyle\sum_{(i,j) \in E} W_{ij}\Big(|x_i - x_j| + \alpha(1 -  \delta(x_i, x_j))\Big)=
    &\displaystyle\sum_{(i,j) \in E} W_{ij}\Big((x_i + x_j -2x_ix_j) + \alpha(1 - (x_i + x_j -1)^2)\Big),
    \label{eq:loss_QUBO}
\end{eqnarray}
where $\delta$ denotes the Kronecker delta and $\alpha$ is a hyperparameter controlling the importance of the smoothness term compared to the min-cut term. Note that minimizing Eq. \ref{eq:loss_QUBO} is equivalent to solving a QUBO problem in Eq. \ref{eq:QUBO}. In particular,
\begin{equation}
\left\{
\begin{array}{ll}
    \mathcal{L}(\vec{x}) =& \vec{x}^T Q \vec{x} + \vec{c}^T \vec{x}, \\
    c_i =& (2\alpha +1)\sum_j W_{ij}, \\
    Q_{ij} =& -2(1+\alpha)W_{ij}, \\
    Q_{ii} =& -\alpha\sum_{j}W_{ij}.
\end{array}
\right.
\end{equation}

In this study, we will investigate various quantum computing and quantum-inspired techniques to address this QUBO problem, and compare their performance against classical benchmarks. The subsequent sections will introduce all the classical and quantum methods employed for the task of image segmentation.

\subsubsection{Simulated annealing}

Simulated annealing is a quantum-inspired technique employed for heuristic optimization of complex models, such as binary quadratic models like QUBO. The objective of simulated annealing is to locate a low-energy state of a system, where the energy is defined by the loss function in Eq. \ref{eq:loss_QUBO}, corresponding to an optimal or near-optimal solution to the optimization problem.

The simulated annealing algorithm iteratively updates the state of the system through a sequence of decreasing temperatures. The state of each variable in the system ($\vec{x}$) is updated according to the Metropolis-Hastings algorithm \citep{Hastings1970}. This process involves proposing a new state $\vec{x'}$ and determining whether to accept it based on the Boltzmann distribution:
\begin{equation}
P(\vec{x} \rightarrow \vec{x'}) = \min(1, e^{-\Delta E \beta} ),
\end{equation}
where $\Delta E = E(\vec{x'}) - E(\vec{x}) = \mathcal{L}(\vec{x'}) - \mathcal{L}(\vec{x})$, and $\beta$ represents the inverse temperature, defined as $\beta = \frac{1}{k_B T}$, with $k_B$ being the Boltzmann constant. As $\beta$ increases, the probability of accepting a higher-energy state decreases, allowing the system to escape local minima and approach a global minimum. This mechanism enables the system to explore a wide range of states at low $\beta$ values and gradually focus on lower-energy states as $\beta$ increases. In this study, we employ a linear scheduler for $\beta$ to control the cooling schedule, linearly increasing $\beta$ within the [0.1, 4.2] range. This linear schedule ensures a smooth transition from exploration to exploitation, effectively balancing the need to escape local minima and converge to a global minimum.

\subsubsection{Quantum annealing}

Quantum annealing is a quantum computing technique designed for solving QUBO problems. The goal is to find the lowest-energy state of a quantum system that encodes the solution to the problem, where the energy corresponds to the objective function of the QUBO problem. Contrary to simulated annealing, quantum annealing leverages quantum tunneling, allowing the system to transition between states by tunneling through energy barriers rather than climbing over them. This capability enables the system to escape local minima more efficiently than classical methods, thus improving the likelihood of finding the global minimum.

The process begins by encoding the QUBO problem into a problem Hamiltonian $H_P$. This involves mapping the QUBO variables to the topology of physical qubits, whose connectivity is constrained by the available working graph of the quantum processor. The system's Hamiltonian evolves from an initial Hamiltonian $H_0$ with a known, easily prepared ground state (a superposition of all possible states) to the problem Hamiltonian $H_P$, whose ground state represents the optimal solution to the QUBO problem. The evolution of the Hamiltonian in quantum annealing is expressed as
\begin{equation}
    H(t) = (1 - s(t)) H_0 + s(t) H_P,
\end{equation}
where $H_0$ is the initial Hamiltonian and $H_P$ is the problem Hamiltonian. The function $s(t)$ varies from 0 to 1 over the annealing time $t$, gradually transforming the system from the ground state of $H_0$ to the ground state of $H_P$. The annealing schedule, defined by the function $s(t)$, is critical to the success of the quantum annealing process. A carefully designed annealing schedule ensures that the system remains in or near the ground state throughout the evolution, thereby increasing the likelihood of reaching the optimal solution.

In this work, we will run our experiments on the D-Wave Sampler with the Pegasus topology, performing 2000 runs. The Pegasus topology is well-suited for quantum annealing due to its structure, which provides enhanced connectivity compared to previous architectures. Moreover, the image segmentation problem described in this work is characterized by having low connectivity, where each node is connected to only four neighboring nodes, making it particularly well-suited for quantum annealing where qubits have limited connectivity.

\subsubsection{Variational quantum circuits}
A promising hybrid approach to solve optimization problems in image segmentation consists of using a VQA. In this framework, a parameterized quantum circuit encodes the binary vector solution $\vec{x}$ of the QUBO problem defined in Eq. \ref{eq:loss_QUBO}. The parameters of the circuit are optimized using classical computing methods, such as gradient-based techniques, to minimize the loss function. VQAs offer more flexibility than quantum annealing, as the loss function to be minimized is not restricted to a QUBO form and can take various forms.

Different encoding methods and variational circuit designs have been explored for VQAs \citep{VQA2,VQA4}, and in particular for image segmentation \citep{VQA3}. In this work, we adopt an amplitude encoding method that requires only $n = \log_2 |V| + 1$ qubits, where $|V|$ is the number of variables. Here, the first $\log_2 |V|$ qubits are used to encode the binary vector, and one ancilla qubit is used to retrieve the probability of each variable being 0 or 1. This encoding method is similar to the Flexible Representation of Quantum Images (FRQI) \citep{FRQI}, widely used to efficiently represent images as quantum states. The parameterized state of the variational circuit in the amplitude encoding scheme is given by
\begin{equation}
    |\psi(\vec{\theta})\rangle = \sum_{i=0}^{n-1} \gamma_i(\vec{\theta})(\alpha_i(\vec{\theta}) |0\rangle_a + \beta_i(\vec{\theta}) |1\rangle_a) \otimes |i\rangle_r
\end{equation}
where the subscript $a$ represents the single ancilla qubit and the subscript $r$ represents the register qubits. The state $\ket{i}_r$ denotes the $i$-th computational basis state and represents one of the QUBO variables $x_i$, or equivalently, one of the pixels of the image. The ancilla qubit $a$ indicates whether the variable $x_i$ should be 0 or 1. By measuring the final state in the computational basis, we obtain $P(x_i=1) = |\beta_i(\vec{\theta})|^2$. With this definition, it holds that $|\alpha_i(\vec{\theta})|^2 + |\beta_i(\vec{\theta})|^2 = 1 \ \forall i$. The optimization process involves minimizing the loss function in Eq. \ref{eq:loss_QUBO}, replacing the variables $x_i$ with their associated probabilities in the current parameterized state:
\begin{eqnarray}
    \mathcal{L}(|\psi(\vec{\theta})\rangle ) =  
    \displaystyle\sum_{(i,j) \in E} W_{ij}\Big((|\beta_i(\vec{\theta})|^2 + |\beta_j(\vec{\theta})|^2 -2|\beta_i(\vec{\theta})|^2|\beta_j(\vec{\theta})|^2) + \alpha (1 - (|\beta_i(\vec{\theta})|^2 + |\beta_j(\vec{\theta})|^2 -1)^2)\Big).
    \label{eq:loss}
\end{eqnarray}
Once the parameters have been optimized, the final state is retrieved by setting $x_i = 1$ if $|\beta_i(\vec{\theta})|^2>0.5$, and $x_i=0$ otherwise.

\begin{figure*}[!ht]
    \centering
    \includegraphics[width=0.85\linewidth]{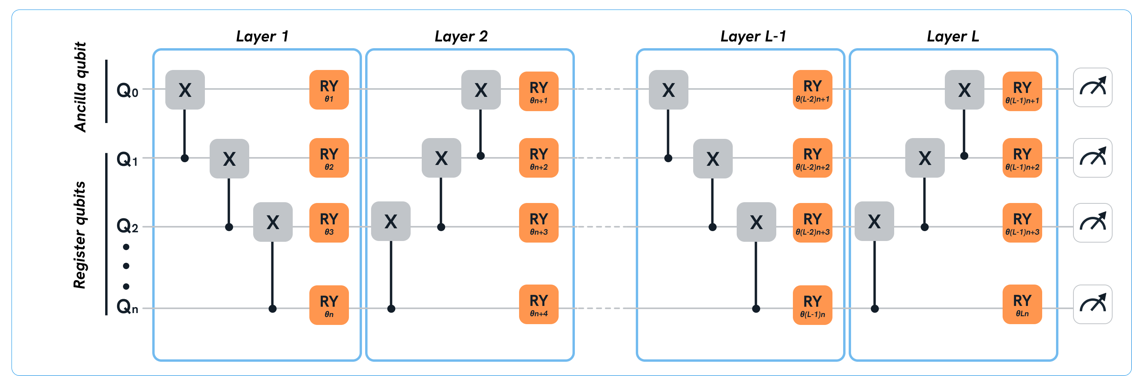}
    \caption{Variational quantum circuit design.}
    \label{fig:circuit}
\end{figure*}

Regarding the ansatz, we use a hardware-efficient ansatz that contains $L$ layers of CNOT gates with linear entanglement and single-qubit rotations $R_y(\theta_k)$. This ansatz, shown schematically in Fig. \ref{fig:circuit}, allows for warm-start optimization. 
Warm-start optimization involves initializing the parameters of the VQA with a heuristic initial guess $\vec{x}^*$ and then optimizing the parameters to refine the solution. This potentially accelerates the optimization process \citep{VQA2}, allowing for faster convergence to the optimal image segmentation. Given an initial solution vector $\vec{x}^*$, the initial state $|\psi_0\rangle$ is calculated by setting $\beta_i = 0$ if $x_i = 0$ and $\beta_i = 1$ if $x_i = 1$ for all variables $x_i$. This leads to the initial state
\begin{equation}
    |\psi_0\rangle = \frac{1}{\sqrt{n}} \sum_{i=0}^{n-1} ((1-x_i)\ket{0}_a + x_i\ket{1}_a) \otimes \ket{i}_r.
\end{equation}
Notice that this state can be easily encoded in the quantum circuit just by combining X and Hadamard gates. In this scenario, the initial state is selected such that $x_i=1$ if the intensity of pixel $i$ in the quantum-inspired transformation from Sect. \ref{sect:Image_transformation} exceeds a certain threshold $T$, which is set to $T=0.3$ in this case.

Finally, we fine-tune this initial state to minimize the loss function. To gradually change the state, we select an ansatz that leads to the identity transformation when the parameters $\vec{\theta}$ are all initialized to zero. This is the case for the ansatz shown in Fig \ref{fig:circuit}. 

The experiments were conducted using the Qibo package to classically simulate the quantum circuit. We used the Adam classical optimizer with a learning rate of $\eta = 0.01$, and the training was performed for 100 epochs. This corresponds to a quantum-inspired algorithm, as the experiments were conducted using classical CPUs. However, this algorithm could also be implemented on gate-based quantum computers.

\subsubsection{Classical methods}
\label{sect:classical-methods}
In this section, we outline the traditional methods employed to benchmark quantum and quantum-inspired techniques. The first two methods, UNET and ResUNET, are supervised neural network-based approaches. The Otsu thresholding method represents a straightforward unsupervised data processing technique, whereas the Gurobi optimization method serves as a classical solver for the QUBO problem detailed in Sect.~\ref{sect:Image_segmentation}. 

\begin{itemize}
    \item \textbf{UNet:} Traditional neural network methods, such as convolutional or fully connected networks, often struggle with low segmentation accuracy, particularly when dealing with small objects and irregular shapes. These limitations arise primarily due to the loss of spatial resolution during the processing of images. To address these challenges, Ronneberger et al. \citep{UNET} introduced a novel model known as UNet. The model derives its name from its distinctive U-shaped architecture, which is composed of a contracting path followed by an expanding path.

The contracting path of the UNet model is responsible for extracting features from the input image and progressively reducing its resolution. This is achieved through a series of convolutional and pooling layers that capture the essential characteristics of the image at various scales. Conversely, the expanding path aims to restore the original size of the image while producing the segmentation map. This path utilizes upsampling operations and convolutional layers to reconstruct the image, integrating both coarse and fine details.

A critical aspect of the UNet architecture is the incorporation of skip connections. These connections link corresponding layers of the contracting and expanding paths, allowing the model to fuse high-resolution features from the contracting path with the upsampled output of the expanding path. This fusion preserves spatial information and enhances the model's ability to learn detailed and accurate segmentations.

In this work, we consider two UNet architectures. The first is a larger model that processes images at their original size of $256 \times 256$ pixels, while the second handles smaller $42 \times 42$ images, allowing for comparison with quantum and quantum-inspired methods. The larger UNet features a contracting path with five convolutional blocks, using filter sizes of 32, 64, 128, 256, and 512, each with a kernel size of 3, ReLU activations, and batch normalization, followed by max-pooling layers that halve the spatial dimensions. Its expanding path mirrors this structure with transposed convolutions for upsampling and corresponding convolutional blocks, also employing ReLU activations and batch normalization, and utilizing skip connections to fuse features from the contracting path. The smaller UNet retains this architecture but comprises only three convolutional blocks with filter sizes of 32, 64, and 128.

    \item  \textbf{ResUNet:}
    The ResUNet model enhances the traditional UNet architecture by replacing standard convolutional blocks with residual convolutional blocks \citep{ResUNET}. These blocks are designed to address the vanishing gradient problem and improve the training efficiency of deep neural networks. They introduce skip connections that allow the input to bypass one or more layers and be directly added to the output of those layers. This creates a direct path for gradients to flow during backpropagation, thereby enhancing the network's learning ability and convergence speed. In this work, we design two ResUNet models to process both large and small images, using the same filter and kernel sizes as the UNet models.
    \item \textbf{Otsu thresholding:} Otsu thresholding is an unsupervised image segmentation technique designed to automatically determine the optimal threshold for separating an image into foreground and background regions. Developed by Nobuyuki Otsu \citep{Otsu}, this method minimizes intra-class variance within the background and foreground regions. Otsu's algorithm iteratively searches for the threshold that minimizes the weighted sum of variances of these two classes. This optimal threshold is then used to binarize the image, assigning pixels with intensities below the threshold to the background and those above it to the foreground.
    \item \textbf{Gurobi optimization:} Gurobi is a classical leading optimization software extensively used for solving a wide range of optimization problems, including QUBOs \citep{gurobi}. It is renowned for its speed and reliability in addressing classical optimization problems.
\end{itemize}

\section{Data Availability}
The dataset used in this study is the INbreast mammography dataset developed in \cite{moreira2012inbreast}, obtained from the corresponding author Inês Domingues, Porto, Portugal, after signing a transfer agreement.

\section{Code Availability}
The code used in this work can be found at  \href{https://github.com/ingenii-solutions/ingenii-quantum-hybrid-networks/blob/main/ingenii_quantum/examples/quantum_inspired_filter.ipynb}{Ingenii's open access library} and Ingenii's \href{https://www.ingenii.io/qml-fundamentals}{Quantum Machine Learning Fundamentals course}.

\bibliography{medical}  

\begin{thebibliography}{10}
\urlstyle{rm}
\expandafter\ifx\csname url\endcsname\relax
  \def\url#1{\texttt{#1}}\fi
\expandafter\ifx\csname urlprefix\endcsname\relax\def\urlprefix{URL }\fi
\expandafter\ifx\csname doiprefix\endcsname\relax\def\doiprefix{DOI: }\fi
\providecommand{\bibinfo}[2]{#2}
\providecommand{\eprint}[2][]{\url{#2}}

\bibitem{ahmad2023diagnosis}
\bibinfo{author}{Ahmad, S.}, \bibinfo{author}{Asghar, M.~Z.}, \bibinfo{author}{Alotaibi, F.~M.} \& \bibinfo{author}{Alotaibi, Y.~D.}
\newblock \bibinfo{journal}{\bibinfo{title}{Diagnosis of cardiovascular disease using deep learning technique}}.
\newblock {\emph{\JournalTitle{Soft Computing}}} \textbf{\bibinfo{volume}{27}}, \bibinfo{pages}{8971--8990} (\bibinfo{year}{2023}).

\bibitem{hussain2024mind}
\bibinfo{author}{Hussain, I.}, \bibinfo{author}{Nazir, M.~B.} \emph{et~al.}
\newblock \bibinfo{journal}{\bibinfo{title}{Mind matters: Exploring ai, machine learning, and deep learning in neurological health}}.
\newblock {\emph{\JournalTitle{International Journal of Advanced Engineering Technologies and Innovations}}} \textbf{\bibinfo{volume}{1}}, \bibinfo{pages}{209--230} (\bibinfo{year}{2024}).

\bibitem{tsietso2023multi}
\bibinfo{author}{Tsietso, D.} \emph{et~al.}
\newblock \bibinfo{journal}{\bibinfo{title}{Multi-input deep learning approach for breast cancer screening using thermal infrared imaging and clinical data}}.
\newblock {\emph{\JournalTitle{IEEE Access}}} \textbf{\bibinfo{volume}{11}}, \bibinfo{pages}{52101--52116} (\bibinfo{year}{2023}).

\bibitem{hendrix2023musculoskeletal}
\bibinfo{author}{Hendrix, N.} \emph{et~al.}
\newblock \bibinfo{journal}{\bibinfo{title}{Musculoskeletal radiologist-level performance by using deep learning for detection of scaphoid fractures on conventional multi-view radiographs of hand and wrist}}.
\newblock {\emph{\JournalTitle{European Radiology}}} \textbf{\bibinfo{volume}{33}}, \bibinfo{pages}{1575--1588} (\bibinfo{year}{2023}).

\bibitem{jyothi2023deep}
\bibinfo{author}{Jyothi, P.} \& \bibinfo{author}{Singh, A.~R.}
\newblock \bibinfo{journal}{\bibinfo{title}{Deep learning models and traditional automated techniques for brain tumor segmentation in mri: a review}}.
\newblock {\emph{\JournalTitle{Artificial intelligence review}}} \textbf{\bibinfo{volume}{56}}, \bibinfo{pages}{2923--2969} (\bibinfo{year}{2023}).

\bibitem{siegel2020cancer}
\bibinfo{author}{Siegel, R.~L.}, \bibinfo{author}{Miller, K.~D.} \& \bibinfo{author}{Jemal, A.}
\newblock \bibinfo{journal}{\bibinfo{title}{Cancer statistics, 2020}}.
\newblock {\emph{\JournalTitle{CA: A Cancer Journal for Clinicians}}} \textbf{\bibinfo{volume}{70}}, \bibinfo{pages}{7--30}, \doiprefix\url{10.3322/caac.21590} (\bibinfo{year}{2020}).

\bibitem{who_breast_cancer}
\bibinfo{author}{Organization, W.~H.}
\newblock \bibinfo{title}{Breast cancer}.
\newblock \bibinfo{howpublished}{\url{https://www.who.int/news-room/fact-sheets/detail/breast-cancer}} (\bibinfo{year}{2024}).
\newblock \bibinfo{note}{Accessed: 2024-10-10}.

\bibitem{laubysecretan2015}
\bibinfo{author}{Lauby-Secretan, B.} \emph{et~al.}
\newblock \bibinfo{journal}{\bibinfo{title}{Breast-cancer screening—viewpoint of the iarc working group}}.
\newblock {\emph{\JournalTitle{The New England Journal of Medicine}}} \textbf{\bibinfo{volume}{372}}, \bibinfo{pages}{2353--2358}, \doiprefix\url{10.1056/NEJMsr1504363} (\bibinfo{year}{2015}).

\bibitem{Otsu}
\bibinfo{author}{Otsu, N.}
\newblock \bibinfo{journal}{\bibinfo{title}{A threshold selection method from gray-level histograms}}.
\newblock {\emph{\JournalTitle{IEEE Transactions on Systems, Man, and Cybernetics}}} \textbf{\bibinfo{volume}{9}}, \bibinfo{pages}{62--66}, \doiprefix\url{10.1109/TSMC.1979.4310076} (\bibinfo{year}{1979}).

\bibitem{shapiro2001computer}
\bibinfo{author}{Shapiro, L.~G.} \& \bibinfo{author}{Stockman, G.~C.}
\newblock \emph{\bibinfo{title}{Computer Vision}} (\bibinfo{publisher}{Prentice Hall}, \bibinfo{year}{2001}).

\bibitem{liang2019watershed}
\bibinfo{author}{Liang, Y.} \& \bibinfo{author}{Fu, J.}
\newblock \bibinfo{journal}{\bibinfo{title}{Watershed algorithm for medical image segmentation based on morphology and total variation model}}.
\newblock {\emph{\JournalTitle{International Journal of Pattern Recognition and Artificial Intelligence}}} \textbf{\bibinfo{volume}{33}}, \bibinfo{pages}{1954019} (\bibinfo{year}{2019}).

\bibitem{meng2021graph}
\bibinfo{author}{Meng, Y.} \emph{et~al.}
\newblock \bibinfo{journal}{\bibinfo{title}{Graph-based region and boundary aggregation for biomedical image segmentation}}.
\newblock {\emph{\JournalTitle{IEEE transactions on medical imaging}}} \textbf{\bibinfo{volume}{41}}, \bibinfo{pages}{690--701} (\bibinfo{year}{2021}).

\bibitem{wang2022medical}
\bibinfo{author}{Wang, R.} \emph{et~al.}
\newblock \bibinfo{journal}{\bibinfo{title}{Medical image segmentation using deep learning: A survey}}.
\newblock {\emph{\JournalTitle{IET image processing}}} \textbf{\bibinfo{volume}{16}}, \bibinfo{pages}{1243--1267} (\bibinfo{year}{2022}).

\bibitem{long2015fully}
\bibinfo{author}{Long, J.}, \bibinfo{author}{Shelhamer, E.} \& \bibinfo{author}{Darrell, T.}
\newblock \bibinfo{title}{Fully convolutional networks for semantic segmentation}.
\newblock In \emph{\bibinfo{booktitle}{Proceedings of the IEEE conference on computer vision and pattern recognition}}, \bibinfo{pages}{3431--3440} (\bibinfo{year}{2015}).

\bibitem{zhu2018adversarial}
\bibinfo{author}{Zhu, W.}, \bibinfo{author}{Xiang, X.}, \bibinfo{author}{Tran, T.~D.}, \bibinfo{author}{Hager, G.~D.} \& \bibinfo{author}{Xie, X.}
\newblock \bibinfo{title}{Adversarial deep structured nets for mass segmentation from mammograms}.
\newblock In \emph{\bibinfo{booktitle}{2018 IEEE 15th International Symposium on Biomedical Imaging (ISBI 2018)}}, \bibinfo{pages}{847--850} (\bibinfo{publisher}{IEEE}, \bibinfo{year}{2018}).

\bibitem{xu2022medical}
\bibinfo{author}{Xu, Y.} \emph{et~al.}
\newblock \bibinfo{journal}{\bibinfo{title}{A medical image segmentation method based on multi-dimensional statistical features}}.
\newblock {\emph{\JournalTitle{Frontiers in Neuroscience}}} \textbf{\bibinfo{volume}{16}}, \bibinfo{pages}{1009581} (\bibinfo{year}{2022}).

\bibitem{UNET}
\bibinfo{author}{Ronneberger, O.}, \bibinfo{author}{Fischer, P.} \& \bibinfo{author}{Brox, T.}
\newblock \bibinfo{title}{U-net: Convolutional networks for biomedical image segmentation}.
\newblock In \emph{\bibinfo{booktitle}{Medical Image Computing and Computer-Assisted Intervention (MICCAI)}}, vol. \bibinfo{volume}{9351}, \bibinfo{pages}{234--241}, \doiprefix\url{10.1007/978-3-319-24574-4_28} (\bibinfo{publisher}{Springer, Cham}, \bibinfo{year}{2015}).

\bibitem{tang2019efficient}
\bibinfo{author}{Tang, P.} \emph{et~al.}
\newblock \bibinfo{journal}{\bibinfo{title}{Efficient skin lesion segmentation using separable-unet with stochastic weight averaging}}.
\newblock {\emph{\JournalTitle{Computer Methods and Programs in Biomedicine}}} \textbf{\bibinfo{volume}{178}}, \bibinfo{pages}{289--301}, \doiprefix\url{10.1016/j.cmpb.2019.07.005} (\bibinfo{year}{2019}).

\bibitem{li2019cascade}
\bibinfo{author}{Li, S.}, \bibinfo{author}{Chen, Y.}, \bibinfo{author}{Yang, S.} \& \bibinfo{author}{Luo, W.}
\newblock \bibinfo{title}{Cascade dense-unet for prostate segmentation in mr images}.
\newblock In \emph{\bibinfo{booktitle}{International Conference on Intelligent Computing}}, \bibinfo{pages}{481--490}, \doiprefix\url{10.1007/978-3-030-26763-6_46} (\bibinfo{publisher}{Springer, Cham}, \bibinfo{year}{2019}).

\bibitem{ConnectedUNets}
\bibinfo{author}{Baccouche, A.} \emph{et~al.}
\newblock \bibinfo{journal}{\bibinfo{title}{Connected-unets: a deep learning architecture for breast mass segmentation}}.
\newblock {\emph{\JournalTitle{npj Breast Cancer}}} \textbf{\bibinfo{volume}{7}}, \bibinfo{pages}{151}, \doiprefix\url{10.1038/s41523-021-00358-x} (\bibinfo{year}{2021}).

\bibitem{soulami2021breast}
\bibinfo{author}{Soulami, K.~B.}, \bibinfo{author}{Kaabouch, N.}, \bibinfo{author}{Saidi, M.~N.} \& \bibinfo{author}{Tamtaoui, A.}
\newblock \bibinfo{journal}{\bibinfo{title}{Breast cancer: one-stage automated detection, segmentation, and classification of digital mammograms using unet model based-semantic segmentation}}.
\newblock {\emph{\JournalTitle{Biomedical Signal Processing and Control}}} \textbf{\bibinfo{volume}{66}}, \bibinfo{pages}{102481}, \doiprefix\url{10.1016/j.bspc.2021.102481} (\bibinfo{year}{2021}).

\bibitem{tariq2023multilevel}
\bibinfo{author}{Tariq~Jamal, A.}, \bibinfo{author}{Abdel-Khalek, S.} \& \bibinfo{author}{Ben~Ishak, A.}
\newblock \bibinfo{journal}{\bibinfo{title}{Multilevel segmentation of medical images in the framework of quantum and classical techniques}}.
\newblock {\emph{\JournalTitle{Multimedia Tools and Applications}}} \textbf{\bibinfo{volume}{82}}, \bibinfo{pages}{13167--13180} (\bibinfo{year}{2023}).

\bibitem{gurobi}
\bibinfo{author}{{Gurobi Optimization, LLC}}.
\newblock \bibinfo{title}{{Gurobi Optimizer Reference Manual}} (\bibinfo{year}{2024}).

\bibitem{dhanachandra2015image}
\bibinfo{author}{Dhanachandra, N.}, \bibinfo{author}{Manglem, K.} \& \bibinfo{author}{Chanu, Y.~J.}
\newblock \bibinfo{journal}{\bibinfo{title}{Image segmentation using k-means clustering algorithm and subtractive clustering algorithm}}.
\newblock {\emph{\JournalTitle{Procedia Computer Science}}} \textbf{\bibinfo{volume}{54}}, \bibinfo{pages}{764--771} (\bibinfo{year}{2015}).

\bibitem{schwab2024deepgaussianmixturemodel}
\bibinfo{author}{Schwab, M.}, \bibinfo{author}{Mayr, A.} \& \bibinfo{author}{Haltmeier, M.}
\newblock \bibinfo{title}{Deep gaussian mixture model for unsupervised image segmentation} (\bibinfo{year}{2024}).
\newblock \eprint{2404.12252}.

\bibitem{PredictiveModels}
\bibinfo{author}{Hothem, D.}, \bibinfo{author}{Hines, J.}, \bibinfo{author}{Nataraj, K.}, \bibinfo{author}{Blume-Kohout, R.} \& \bibinfo{author}{Proctor, T.}
\newblock \bibinfo{title}{Predictive models from quantum computer benchmarks}.
\newblock In \emph{\bibinfo{booktitle}{2023 IEEE International Conference on Quantum Computing and Engineering (QCE)}}, vol.~\bibinfo{volume}{01}, \bibinfo{pages}{709--714}, \doiprefix\url{10.1109/QCE57702.2023.00086} (\bibinfo{year}{2023}).

\bibitem{Konar2020QI}
\bibinfo{author}{Konar, D.}, \bibinfo{author}{Bhattacharyya, S.}, \bibinfo{author}{Gandhi, T.~K.} \& \bibinfo{author}{Panigrahi, B.~K.}
\newblock \bibinfo{journal}{\bibinfo{title}{A quantum-inspired self-supervised network model for automatic segmentation of brain mr images}}.
\newblock {\emph{\JournalTitle{Applied Soft Computing}}} \textbf{\bibinfo{volume}{93}}, \bibinfo{pages}{106348}, \doiprefix\url{10.1016/j.asoc.2020.106348} (\bibinfo{year}{2020}).

\bibitem{Konar2023TN}
\bibinfo{author}{Konar, D.}, \bibinfo{author}{Bhattacharyya, S.}, \bibinfo{author}{Gandhi, T.~K.} \& \bibinfo{author}{et~al.}
\newblock \bibinfo{journal}{\bibinfo{title}{3d quantum-inspired self-supervised tensor network for volumetric segmentation of brain mr images}}.
\newblock {\emph{\JournalTitle{TechRxiv}}} \doiprefix\url{10.36227/techrxiv.12909860.v4} (\bibinfo{year}{2023}).

\bibitem{QAradar}
\bibinfo{author}{Presles, T.} \emph{et~al.}
\newblock \bibinfo{journal}{\bibinfo{title}{Synthetic aperture radar image segmentation with quantum annealing}}.
\newblock {\emph{\JournalTitle{IET Radar, Sonar \& Navigation}}} \textbf{\bibinfo{volume}{18}}, \bibinfo{pages}{812--824}, \doiprefix\url{10.1049/rsn2.12523} (\bibinfo{year}{2024}).

\bibitem{venkatesh2024q}
\bibinfo{author}{Venkatesh, S.~M.}, \bibinfo{author}{Macaluso, A.}, \bibinfo{author}{Nuske, M.}, \bibinfo{author}{Klusch, M.} \& \bibinfo{author}{Dengel, A.}
\newblock \bibinfo{journal}{\bibinfo{title}{Q-seg: Quantum annealing-based unsupervised image segmentation}}.
\newblock {\emph{\JournalTitle{IEEE Computer Graphics and Applications}}}  (\bibinfo{year}{2024}).

\bibitem{wang2024implementation}
\bibinfo{author}{Wang, K.} \emph{et~al.}
\newblock \bibinfo{journal}{\bibinfo{title}{Implementation and analysis of quantum-classical hybrid interactive image segmentation algorithm based on quantum annealer}}.
\newblock {\emph{\JournalTitle{Quantum Information Processing}}} \textbf{\bibinfo{volume}{23}}, \bibinfo{pages}{301} (\bibinfo{year}{2024}).

\bibitem{Inbreast}
\bibinfo{author}{Moreira, I.~C.} \emph{et~al.}
\newblock \bibinfo{journal}{\bibinfo{title}{Inbreast: toward a full-field digital mammographic database}}.
\newblock {\emph{\JournalTitle{Academic Radiology}}} \textbf{\bibinfo{volume}{19}}, \bibinfo{pages}{236--248}, \doiprefix\url{10.1016/j.acra.2011.09.014} (\bibinfo{year}{2012}).

\bibitem{Konar2023QuantumTensor}
\bibinfo{author}{Konar, D.}, \bibinfo{author}{Bhattacharyya, S.}, \bibinfo{author}{Gandhi, T.~K.} \emph{et~al.}
\newblock \bibinfo{journal}{\bibinfo{title}{3d quantum-inspired self-supervised tensor network for volumetric segmentation of brain mr images}}.
\newblock {\emph{\JournalTitle{TechRxiv}}} \doiprefix\url{10.36227/techrxiv.12909860.v4} (\bibinfo{year}{2023}).
\newblock \bibinfo{note}{Preprint}.

\bibitem{CompactifAI}
\bibinfo{author}{Mugel, S.} \& \bibinfo{author}{Or{\'u}s, R.}
\newblock \bibinfo{journal}{\bibinfo{title}{Compactifai: Extreme compression of large language models}}.
\newblock {\emph{\JournalTitle{arXiv:2401.14109}}}  (\bibinfo{year}{2024}).

\bibitem{CP}
\bibinfo{author}{Lebedev, V.}, \bibinfo{author}{Ganin, Y.}, \bibinfo{author}{Rakhuba, M.}, \bibinfo{author}{Oseledets, I.} \& \bibinfo{author}{Lempitsky, V.}
\newblock \bibinfo{journal}{\bibinfo{title}{Speeding-up convolutional neural networks using fine-tuned cp-decomposition}}.
\newblock {\emph{\JournalTitle{arXiv:1511.06530}}}  (\bibinfo{year}{2015}).

\bibitem{tucker}
\bibinfo{author}{Kim, Y.-D.} \emph{et~al.}
\newblock \bibinfo{journal}{\bibinfo{title}{Compression of deep convolutional neural networks for fast and low power mobile applications}}.
\newblock {\emph{\JournalTitle{arXiv:1511.06530}}}  (\bibinfo{year}{2015}).

\bibitem{qing2024compressing}
\bibinfo{author}{Qing, Y.}, \bibinfo{author}{Li, K.}, \bibinfo{author}{Zhou, P.-F.} \& \bibinfo{author}{Ran, S.-J.}
\newblock \bibinfo{journal}{\bibinfo{title}{Compressing neural network by tensor network with exponentially fewer variational parameters}}.
\newblock {\emph{\JournalTitle{arXiv:2305.06058}}}  (\bibinfo{year}{2024}).

\bibitem{YOLO}
\bibinfo{author}{Baccouche, A.}, \bibinfo{author}{Garcia-Zapirain, B.}, \bibinfo{author}{Castillo~Olea, C.} \& \bibinfo{author}{Elmaghraby, A.~S.}
\newblock \bibinfo{journal}{\bibinfo{title}{Breast lesions detection and classification via yolo-based fusion models}}.
\newblock {\emph{\JournalTitle{Computational Materials Continuum}}} \textbf{\bibinfo{volume}{69}}, \bibinfo{pages}{1407--1425} (\bibinfo{year}{2021}).

\bibitem{Histogram}
\bibinfo{author}{Pizer, S.}, \bibinfo{author}{Johnston, R.}, \bibinfo{author}{Ericksen, J.}, \bibinfo{author}{Yankaskas, B.} \& \bibinfo{author}{Muller, K.}
\newblock \bibinfo{title}{Contrast-limited adaptive histogram equalization: speed and effectiveness}.
\newblock In \emph{\bibinfo{booktitle}{[1990] Proceedings of the First Conference on Visualization in Biomedical Computing}}, \bibinfo{pages}{337--345}, \doiprefix\url{10.1109/VBC.1990.109340} (\bibinfo{year}{1990}).

\bibitem{Bhattacharyya2011}
\bibinfo{author}{Bhattacharyya, S.}, \bibinfo{author}{Maulik, U.} \& \bibinfo{author}{Dutta, P.}
\newblock \bibinfo{journal}{\bibinfo{title}{Multilevel image segmentation with adaptive image context based thresholding}}.
\newblock {\emph{\JournalTitle{Applied Soft Computing}}} \textbf{\bibinfo{volume}{11}}, \bibinfo{pages}{946--962}, \doiprefix\url{10.1016/j.asoc.2010.01.015} (\bibinfo{year}{2011}).

\bibitem{QA}
\bibinfo{author}{Venkatesh, S.~M.}, \bibinfo{author}{Macaluso, A.}, \bibinfo{author}{Nuske, M.}, \bibinfo{author}{Klusch, M.} \& \bibinfo{author}{Dengel, A.}
\newblock \bibinfo{title}{Q-seg: Quantum annealing-based unsupervised image segmentation} (\bibinfo{year}{2023}).
\newblock \eprint{2311.12912}.

\bibitem{venkatesh2}
\bibinfo{author}{Venkatesh, S.~M.}, \bibinfo{author}{Macaluso, A.}, \bibinfo{author}{Nuske, M.}, \bibinfo{author}{Klusch, M.} \& \bibinfo{author}{Dengel, A.}
\newblock \bibinfo{journal}{\bibinfo{title}{Qubit-efficient variational quantum algorithms for image segmentation}}.
\newblock {\emph{\JournalTitle{arXiv preprint arXiv:2405.14405}}}  (\bibinfo{year}{2024}).

\bibitem{lucas2014ising}
\bibinfo{author}{Lucas, A.}
\newblock \bibinfo{journal}{\bibinfo{title}{Ising formulations of many np problems}}.
\newblock {\emph{\JournalTitle{Frontiers in physics}}} \textbf{\bibinfo{volume}{2}}, \bibinfo{pages}{5} (\bibinfo{year}{2014}).

\bibitem{mcgeoch2020theory}
\bibinfo{author}{McGeoch, C.~C.}
\newblock \bibinfo{journal}{\bibinfo{title}{Theory versus practice in annealing-based quantum computing}}.
\newblock {\emph{\JournalTitle{Theoretical Computer Science}}} \textbf{\bibinfo{volume}{816}}, \bibinfo{pages}{169--183} (\bibinfo{year}{2020}).

\bibitem{Potts}
\bibinfo{author}{Boykov, Y.}, \bibinfo{author}{Veksler, O.} \& \bibinfo{author}{Zabih, R.}
\newblock \bibinfo{title}{Fast approximate energy minimization via graph cuts}.
\newblock In \emph{\bibinfo{booktitle}{IEEE Transactions on Pattern Analysis and Machine Intelligence}}, vol.~\bibinfo{volume}{23}, \bibinfo{pages}{1222--1239}, \doiprefix\url{10.1109/34.969114} (\bibinfo{year}{2001}).

\bibitem{Hastings1970}
\bibinfo{author}{Hastings, W.~K.}
\newblock \bibinfo{journal}{\bibinfo{title}{Monte carlo sampling methods using markov chains and their applications}}.
\newblock {\emph{\JournalTitle{Biometrika}}} \textbf{\bibinfo{volume}{57}}, \bibinfo{pages}{97--109}, \doiprefix\url{10.1093/biomet/57.1.97} (\bibinfo{year}{1970}).

\bibitem{VQA2}
\bibinfo{author}{Perelshtein, M.~R.} \emph{et~al.}
\newblock \bibinfo{journal}{\bibinfo{title}{{NISQ-compatible approximate quantum algorithm for unconstrained and constrained discrete optimization}}}.
\newblock {\emph{\JournalTitle{Quantum}}} \textbf{\bibinfo{volume}{7}}, \bibinfo{pages}{1186} (\bibinfo{year}{2023}).

\bibitem{VQA4}
\bibinfo{author}{Venkatesh, S.~M.}, \bibinfo{author}{Macaluso, A.}, \bibinfo{author}{Nuske, M.}, \bibinfo{author}{Klusch, M.} \& \bibinfo{author}{Dengel, A.}
\newblock \bibinfo{title}{Quantum annealing-based algorithm for efficient coalition formation among leo satellites} (\bibinfo{year}{2024}).
\newblock \eprint{2408.06007}.

\bibitem{VQA3}
\bibinfo{author}{Venkatesh, S.~M.}, \bibinfo{author}{Macaluso, A.}, \bibinfo{author}{Nuske, M.}, \bibinfo{author}{Klusch, M.} \& \bibinfo{author}{Dengel, A.}
\newblock \bibinfo{title}{Qubit-efficient variational quantum algorithms for image segmentation} (\bibinfo{year}{2024}).
\newblock \eprint{2405.14405}.

\bibitem{FRQI}
\bibinfo{author}{Le, P.}, \bibinfo{author}{Iliyasu, A.}, \bibinfo{author}{Dong, F.} \& \bibinfo{author}{Hirota, K.}
\newblock \bibinfo{journal}{\bibinfo{title}{A flexible representation of quantum images for polynomial preparation, image compression and processing operations, quantum inf}}.
\newblock {\emph{\JournalTitle{Quantum Information Processing}}} \textbf{\bibinfo{volume}{10}}, \bibinfo{pages}{63--84}, \doiprefix\url{10.1007/s11128-010-0177-y} (\bibinfo{year}{2011}).

\bibitem{ResUNET}
\bibinfo{author}{Zhang, Z.}, \bibinfo{author}{Liu, Q.} \& \bibinfo{author}{Wang, Y.}
\newblock \bibinfo{title}{Road extraction by deep residual u-net}.
\newblock In \emph{\bibinfo{booktitle}{IEEE Geoscience and Remote Sensing Letters}}, vol.~\bibinfo{volume}{15}, \bibinfo{pages}{749--753}, \doiprefix\url{10.1109/LGRS.2018.2802944} (\bibinfo{year}{2018}).

\bibitem{moreira2012inbreast}
\bibinfo{author}{Moreira, I.~C.} \emph{et~al.}
\newblock \bibinfo{journal}{\bibinfo{title}{Inbreast: toward a full-field digital mammographic database}}.
\newblock {\emph{\JournalTitle{Academic radiology}}} \textbf{\bibinfo{volume}{19}}, \bibinfo{pages}{236--248} (\bibinfo{year}{2012}).

\end{thebibliography}

\end{document}